%% file: main.tex
\documentclass{article}

\PassOptionsToPackage{numbers, sort&compress}{natbib}
\usepackage[creativeai, final]{neurips_2025}

\usepackage[utf8]{inputenc} %
\usepackage{hyperref}       %
\usepackage{url}            %
\usepackage{booktabs}       %
\usepackage{amsfonts}       %
\usepackage{nicefrac}       %
\usepackage{microtype}      %
\usepackage{xcolor}         %
\usepackage{amsmath,amssymb}
\usepackage{graphicx}
\usepackage{array}
\usepackage{makecell}
\usepackage{subcaption}
\usepackage[inline]{enumitem}
\usepackage{pifont}
\usepackage{cleveref}
\usepackage{multicol}

\newcommand{\mrtunit}{chunk}

\title{Live Music Models}

\author{%
  Lyria Team, Google DeepMind\footnote{See Contributions and Acknowledgments section for full author list}
}

\begin{document}

\maketitle
\footnotetext[1]{See \hyperref[sec:contributions]{Contributions and Acknowledgments} section for full author list.}

\begin{abstract}

We introduce a new class of generative models for music called 
\emph{live music models} 
that produce a continuous stream of music in real-time with 
synchronized user 
control. 
We release Magenta RealTime, an open-weights live music model that can be steered using text or audio prompts to control acoustic style. 
On automatic metrics of music quality, Magenta RealTime outperforms other open-weights music generation models, 
despite using fewer parameters and offering first-of-its-kind live generation capabilities. 
We also release Lyria RealTime,
an API-based model with extended controls, 
offering access to our most powerful model with wide prompt coverage. 
These models demonstrate a new paradigm for AI-assisted music creation that 
emphasizes 
human-in-the-loop interaction for live music performance.

\textbf{Magenta RealTime} (Open): \url{github.com/magenta/magenta-realtime} \\
\textbf{Lyria RealTime} (API): \url{g.co/magenta/lyria-realtime}

\end{abstract}

\section{Introduction}\label{sec:introduction}

Music exists in two complementary forms: as static recorded pieces (``music as a noun''), and as live performances collectively experienced in real time (``music as a verb'')~\cite{musicking1998small}. 
This second form of \emph{live} music is particularly tied to the fundamental human experiences of creative flow~\cite{flow2013, flow2015}, embodied expression~\cite{Jensenius2022SoundActions}, and social connection~\cite{entrainment2024}. 
Despite this, modern generative AI systems for musical audio
have had an overwhelming emphasis on offline, turn-based generation~\cite{dhariwal2020jukebox,Forsgren_Martiros_2022,Agostinelli2023,copet2023simple,Donahue2023,musicldm2024,evans2024long,evans2024fast,evans2025stable,yuan2025yuescalingopenfoundation}.

Live music represents a new frontier for generative AI, one with numerous opportunities and technical challenges. 
In the conventional offline setting, 
users input control information, 
wait $L$ seconds (offline \emph{latency}), 
and receive $T$ seconds of audio.  
In our proposed live setting, 
users continuously input control information, 
receiving $T$ seconds of an uninterrupted audio stream from $T$ seconds of interaction, 
with $D$ seconds of \emph{delay} between their control inputs and their influence on the audio stream. 
Placing users in a continuous perception-action loop promotes more active creation, creates higher-bandwidth interaction, fosters 
personalized expression and emphasize the process as much as the product. 
However, meeting these rigid synchronization requirements while maintaining high quality audio generation is 
a challenging task for machine learning models. 
Some offline music generation  systems~\cite{novack2025fasttexttoaudiogenerationadversarial} have a Real Time Factor (RTF) 
r$\geq 1\times$, i.e.,~they generate $T$ seconds of audio with latency $L\leq T$.\footnote{RTF is commonly defined as both $L/T$ and $T/L$. Here we use $T/L$, i.e., higher RTF means faster.}
However, most are not well-suited for live performance, as they lack other 
necessary attributes. 
Specifically, we differentiate \emph{live music models} as those which have all three of the following attributes: 
(1)~\emph{real-time generation} with throughput RTF $\geq 1\times$, 
(2)~\emph{causal streaming} 
where audio generates continuously as a function of both user control inputs and past audio output, 
and 
(3)~\emph{responsive controls} (delay $D$ is low, facilitating live interaction).

Open-weights models are particularly well-suited for live generative music because they 
can run locally on users' devices. 
On-device inference  
has numerous benefits in music~\citep{zhou2024local}, enabling 
(1)~\emph{lower latency} by eliminating network requests, 
(2)~\emph{higher reliability}, facilitating usage in real-world contexts, 
(3)~\emph{privacy} guarantees, and
(4)~\emph{customization} by artists. 
Cloud-based APIs address complementary use cases, 
offering access to models running on specialized hardware that are more powerful than those that would run on edge devices,
at the cost of some of the benefits of open-weights models. 

\begin{figure}[t]
    \centering
     \includegraphics[width=\linewidth]{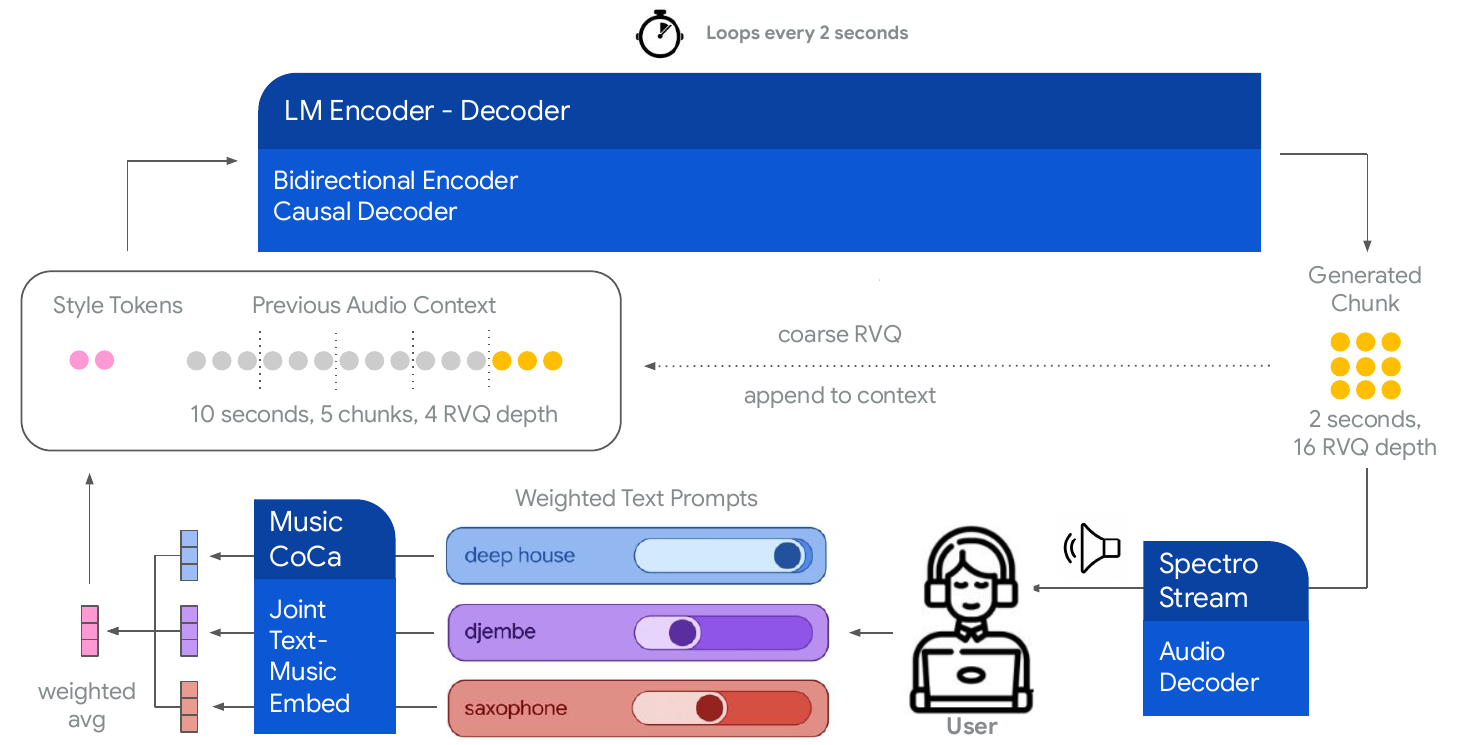}
    \caption{
    Magenta RealTime is a \emph{live music model} that generates an uninterrupted stream of music and responds continuously to user input. 
    It generates audio in two-second \mrtunit{}s using a pipeline with three components:
    (1)~MusicCoCa, a \emph{style embedding} model, 
    (2)~SpectroStream~\citep{li2025spectrostream}, an \emph{audio codec} model, 
    and 
    (3)~an encoder-decoder \emph{language model}. 
    For each \mrtunit,
    a style embedding is computed via a weighted average of MusicCoCa embeddings of text and audio prompts from the user. 
    Given this style embedding and $10$ seconds ($5$ \mrtunit{}s) of past audio context, the language model decoder generates SpectroStream audio tokens for the new \mrtunit{}, which is then decoded to audio. 
    }
    \label{fig:overview}
    \vspace{-5mm}
\end{figure}

To allow users to navigate these tradeoffs based on their application goals, we introduce a pair of systems that span both paradigms: Magenta RT (open-weights, on-device) and Lyria RT (API, cloud-based).
Both use the same core methodological framework, which centers around codec language modeling~\citep{van2017neural,wu2024codeclm} (\Cref{fig:overview}). 
Specifically, we train a language model (LM) to generate audio tokens from SpectroStream~\citep{li2025spectrostream}, using 
a method similar to~\citep{yang2023uniaudio,defossez2024moshispeechtextfoundationmodel} to achieve live streaming~\citep{mcwilliams2025depthformer}.

We focus our subsequent discussion primarily on Magenta RT, 
which may be of higher interest to the AI research community for its ability to be finetuned with new controls~\citep{lin2023content}  
or explored for transfer learning~\citep{castellon2021codified}.

Magenta RT offers first-of-its-kind live generation among open-weights models. 
On automatic metrics of music quality, 
it outperforms existing offline open-weights music generation models like MusicGen (Large)~\citep{copet2023simple} and
Stable Audio Open~\cite{evans2025stable}. 
Moreover, Magenta RT uses 
$38$\% 
fewer parameters ($750$M) than Stable Audio Open ($1.2$B), and 
$77$\% 
fewer than MusicGen Large ($3.3$B). Along with our codebase and model weights, we release a set of demos that run in real time on free-tier Colab TPUs (\texttt{v2-8}) and showcase three distinct use cases: 
live generation, finetuning, and 
a novel live audio input interaction that we call 
\emph{audio injection} (Section~\ref{sec:audio_injection}).\footnote{Audio samples and links to code, weights and demos are provided in the online supplement: \\ \url{https://storage.googleapis.com/live-music-models/index.html}}

\section{Method}\label{sec:method} 

Magenta RT is a codec language model~\citep{van2017neural,wu2024codeclm} designed to generate high-fidelity stereo audio in real time based on acoustic style conditioning. 
To achieve this, we adopt a pipeline-based approach, 
using an encoder-decoder Transformer~\citep{Vaswani2017} to model audio tokens from SpectroStream~\citep{li2025spectrostream} conditioned on style embeddings from our proposed MusicCoCa (\Cref{fig:overview}). See \Cref{sec:related_work} for related work.

\subsection{Audio Tokenization via SpectroStream}
\label{sec:acoustic-tokenizer}

Codec language modeling involves the use of a discrete audio codec to convert audio data into language-like \emph{tokens}. 
A codec is a pair of functions, an encoder and decoder, that convert audio to and from a compressed space with minimal perceivable distortion. 
More formally, the encoder component $\texttt{Enc}$ is a function mapping raw stereo audio waveforms $\mathbf{a} \in \mathbb{R}^{T f_s  \times 2}$ into sequences of discrete tokens ${\mathbb{V}_c^{T f_k \times d_c}}$,
where $T$ is the duration in seconds, $f_s$ the audio sampling rate, $\mathbb{V}_c$ the codec vocabulary, $f_k$ the token frame rate, and $d_c$ the RVQ depth. 
The decoder module $\texttt{Dec} \approxeq \texttt{Enc}^{-1}$ then performs the approximate inverse operation to reconstruct the waveform given the compressed representation, ensuring that the process is as perceptually lossless as possible.

Here we adopt the recently proposed SpectroStream  codec~\cite{li2025spectrostream}, a full-band ($f_s = 48$kHz) multi-channel neural audio codec based on residual vector quantization (RVQ)~\citep{Zeghidour2021}, with an overall bandwidth of $16$kbps ($f_k = 25$Hz, $d_c = 64$, $|\mathbb{V}_c| = 1024$). 
To facilitate live streaming, we reduce the bandwidth to $4$kbps by generating only the first $16$ RVQ levels (coarse and medium from~\Cref{fig:magenta_rt_global}), 
yielding a live throughput target of $400$ tokens per second. 
See~\Cref{sec:spectrostream} for more details.

\subsection{Style Embeddings via MusicCoCa}
\label{sec:musiccoca}

We use a joint audio-text representation as a control mechanism for overall audio style. 
This is achieved by training a joint audio-text embedding model on music annotated with diverse textual descriptions.
The text encapsulates musical characteristics useful for high-level control, which we collectively refer to as \emph{style} (Section~\ref{sec:style_conditioning}).

Our embedding model, MusicCoCa, builds upon MuLan \cite{Huang2022} and CoCa \cite{yu2022coca}.
It is a contrastive captioner (CoCa) consisting of two embedding towers, mapping each modality to a shared $768$-dimensional space. The audio embedding tower $M_A$ is a $12$-layer VisionTransformer (ViT) \cite{dosovitskiy2021an}. Its input is a log-mel spectrogram of a $10$s slice of $16$kHz audio ($128$ channels and length $992$; split into patches of size $16\times16$).
The text embedding tower $M_T$ is a $12$-layer Transformer, which operates on tokenized text with a maximum sequence length of $128$ tokens.
We use attention pooling to reduce the activations of each tower to a single $768$d embedding, which can be subsequently quantized into $12$ discrete tokens with codebook size $|\mathbb{V}_m| = 1024$.
This tokenized representation (of audio or text) is then used as a conditioning signal in the LM encoder.
Variable-length audio of duration $T$ may be embedded by zero padding to a multiple of $10$ seconds and then mean pooling across $\lceil \frac{T}{10} \rceil$ chunks. 

In addition to the two embedding towers, MusicCoCa has a text decoder which can generate audio text captions.
In our application this decoder is a shallow 3-layer Transformer which only serves a regularizing purpose.
The MusicCoCa optimization objective, based on the CoCa framework \cite{yu2022coca}, consists of contrastive and generative loss components which we weigh equally. We train the model using the Adafactor optimizer with a learning rate of $1 \times 10^{-4}$ and $1{,}000$ warmup steps for a total of $16{,}000$ steps. The batch size is $1024$.

\subsection{Modeling Framework}\label{sec:modeling_framework}

Magenta RT operates on audio tokens from a discrete audio codec, 
following an established practice in 
audio modeling \cite{van2017neural,Borsos2022, Agostinelli2023,copet2023simple}. 
Our model autoregressively predicts discrete audio tokens conditioned on both preceding audio and a shared audio-text embedding. 
This modeling mechanism partly mirrors the MusicLM architecture \cite{Agostinelli2023}, itself based on AudioLM \cite{Borsos2022}. 
Similarly to MusicLM, we use a Transformer-based architecture \cite{Vaswani2017} for the token prediction model 
and enable text conditioning by leveraging a joint audio-text embedding model, using the target audio signal at training time, and text inputs at inference time. 
To facilitate live generation, we propose two high-level changes relative to MusicLM: 
(1)~we replace the hierarchical cascade of multiple LMs with a single LM, using a recent method~\citep{mcwilliams2025depthformer} similar to~\citep{yang2023uniaudio,defossez2024moshispeechtextfoundationmodel} for efficiency, and
(2)~we propose a chunk-based autogressive approach to allow for infinite streaming generation.

More formally, given audio $\mathbf{a}$, our goal is to model the probability ${P_\theta(\texttt{Enc}(\mathbf{a}) \mid M_A(\mathbf{a}))}$ of the corresponding sequence of acoustic tokens given its associated style embedding. Here, 
 $\texttt{Enc}$ denotes the audio codec encoder and $M_A$ the MusicCoCa audio encoder.
At inference, we sample from the model $\mathbf{e}\prime \sim P_\theta(\cdot \mid \mathbf{c})$,
conditioning on a style embedding $\mathbf{c}$ obtained from an arbitrary mixture of audio and text prompts.
Finally, we use the codec decoder to produce output audio, i.e.,~$\mathbf{a}\prime = \texttt{Dec}(\mathbf{e}\prime)$.

\subsubsection{Chunk-based autoregression with coarse context} 
\label{sec:chunk-based-autoregression}

\newcommand{\chunk}[1]{\texttt{Chunk}}
\newcommand{\coarse}[1]{\texttt{Coarse}}

In the live generation setting, we require our model to generate an infinite and uninterrupted stream of audio with a RTF $\geq 1\times$. 
To achieve this, we propose two key techniques: chunk-based autoregression to enable infinite streaming, and the use of coarser RVQ tokens in the audio history.

In order to generate an infinite stream of audio, at inference time we must be able to predict an audio sequence with a length likely larger than what the model has seen during training. 
Such a mismatch in sequence length between training and inference is often found to result in unpredictable behavior and degraded performance \cite{Vaswani2017, Huang2018}. Previous work has addressed this issue via sliding attention windows with a relative positional encoding scheme \cite{Su2021,Press2021}, informing the model about the relative distance between tokens instead of describing their absolute position within the sequence. 

We instead propose \emph{chunk-based autoregression}, 
where we operate on chunks of length $C = 2$ seconds, and, under a Markov assumption, predict each chunk based on a limited context of $H = 5$ previous chunks ($10$ seconds of history). 
This has several advantages: it reduces error accumulation and allows for stateless inference, eliminating the need to maintain a generation cache and simplifying model deployment. It also introduces flexibility during sampling, since conditioning is updated between calls without preserving information about controls beyond the context window. 

To achieve RTF $\geq 1\times$, we also propose to use a \emph{coarse representation of the audio context}. 
Due to the hierarchical structure of RVQ, lower quantization levels capture the most salient acoustic information. While generating target tokens at the full RVQ depth ($d_c=16$) is crucial to maintaining high fidelity, 
a lower resolution may be sufficient to represent the audio history.
Therefore, 
we use a coarser representation consisting of the first $4$ RVQ tokens for conditioning on the previous chunks.

More formally, a \emph{chunk} is a contiguous segment of audio tokens representing $C$ seconds of audio. 
For audio $\mathbf{a}$, we define $\chunk{}_i \triangleq \texttt{Enc}(\mathbf{a})_{C f_k i:C f_k (i+1)}$, i.e.,~the span of tokens representing audio between $Ci$ and $C(i+1)$ seconds and including the first $16$ RVQ tokens for each frame. 
We also define $\coarse{}_i$ as the first $4$ RVQ tokens over the same span of time.
Our de facto modeling objective is thus $P_{\theta}(\chunk{}_i \mid 
\coarse{}_{i-H:i}, \mathbf{c}_i)$,
where 
$\mathbf{c}_i = \texttt{Quantize}(M_A(\mathbf{a})_{\lfloor \frac{Ci}{10} \rfloor})$ is $12$ tokens representing the most recent quantized MusicCoCa audio embedding for chunk $i$.

\subsection{Encoder-Decoder Language Model}

To model this distribution, we use an encoder-decoder Transformer~\cite{Vaswani2017} LM trained with T5X~\cite{raffel2020exploring,roberts-etal-23-scaling}. We release pre-trained models using the T5 ~\citep{raffel2020exploring} Base and Large configurations.

\paragraph{Encoder} 
The bidirectional encoder is responsible for processing the acoustic history and style control into an intermediate representation for generation.
At chunk $i$, the encoder receives the concatenation of the acoustic history and style tokens   
${\mathbf{x}_i = \coarse{}_{i-H} \oplus \dots \oplus \coarse{}_{i-1} \oplus \mathbf{c}_i}$,
with a total length of $1012$ tokens ($4\cdot C \cdot H \cdot f_k = 1000$ audio + $12$ style tokens) and a vocabulary unified across the codec and quantized style tokens $\mathbb{V} = \{\texttt{<S>}, \texttt{<P>}\} \cup \mathbb{V}_c \cup \mathbb{V}_m$.

\paragraph{Decoder}
A key differentiating factor compared to prior work is our imposed constraint of achieving RTF $\geq 1\times$ 
generation to enable live interactive applications. 
Past work such as MusicLM~\citep{Agostinelli2023} proposes a hierarchical cascade of language models to model tokens efficiently, while MusicGen~\citep{copet2023simple} proposes a delay pattern---neither approach would achieve RTF $\geq 1\times$ for full bandwidth audio tokens. 
Instead, based on \cite{mcwilliams2025depthformer}, our decoder comprises two connected Transformer modules. The ``temporal'' module constructs a temporal context by processing acoustic frames, where RVQ tokens within each frame are embedded and aggregated to yield a single frame-level embedding. The ``depth'' module then performs autoregressive prediction of the RVQ indices conditioned on the previous temporal context. 
With this setup, we achieve RTF=$1.8$ on H100 GPU with the T5 Large configuration. 

\section{Experiments}
\label{sec:results}
\input{results}

\section{Controllable Generation}
\label{sec:controllable_generation}
\input{controls}

\section{Conclusion}\label{sec:conclusion}

In this work, we introduced live music models, a new class of generative systems designed for real-time, continuous music creation with synchronized user control. We presented two such systems: Magenta RealTime, a fully open-weights model, and Lyria RealTime, an API-based model with extended controls. These models facilitate a novel paradigm for AI-assisted music, emphasizing interactive, human-in-the-loop performance that prioritizes the creative process over just the end product. 
With future work, we aim to further decrease the control latency to unlock new interactive possibilities. Ultra-low latency could enable direct MIDI or audio control, akin to a new class of synthesizer or audio effect. Further, training on multi-stem audio would open the possibility for models to act as musical partners, jamming along with users and providing dynamic live accompaniment.

\newpage

\input{contributions}

\newpage

\input{output.bbl}
\newpage

\appendix

\section{Related Work}
\label{sec:related_work}

Our work on live music models connects to longstanding goals in computer music of live, computer-aided performance. 
Earlier work centered around score following: 
tracking live performances against known musical material to enable computer accompaniment~\citep{dannenberg1984line,vercoe1984synthetic,orio2003score}, or systems that could both track input from human musicians and generate new symbolic material using simple probabilistic models~\citep{lewis2000too,thom2000bob,pachet2003continuator,kondak2016active}.
More recently, 
numerous systems have proposed live interaction with generative AI models of symbolic music~\citep{donahue2019piano,jiang2020rl,benetatos2020bachduet,blanchard2024developing,wu2025adaptive}, 
or small models trained on narrow music audio distributions~\citep{engel2020ddsp,caillon2021rave}. 
Here we aim to bridge the gap between the live interaction paradigm and the broad audio generation and control capabilities of large-scale generative AI models of music audio.

Many offline music audio generation models are based off of codec LMs~\citep{van2017neural,wu2024codeclm}. 
Codec LMs are theoretically capable of streaming provided they meet two criteria: 
(1)~the language model and codec are \emph{causal} (outputs never depend on future inputs), and
(2)~they generate with RTF $\geq 1\times$ for some chunk size $C$ (which lower bounds the control delay $D$). 
To the best of our knowledge, no existing codec LM music generation model~\citep{dieleman2018challenge,dhariwal2020jukebox,Agostinelli2023,copet2023simple,yuan2025yuescalingopenfoundation,liu2025songgen,gong2025ace} satisfies both criteria (most use non-causal codecs). 
Other music generation systems~\citep{Forsgren_Martiros_2022,musicldm2024,evans2024long,evans2025stable,evans2024fast} are based off of latent diffusion---many achieve a throughput RTF $\geq 1\times$ but generate in a non-causal fashion. 
Some ``outpainting'' methods have been proposed for latent diffusion music models~\citep{levy2023controllable,novack2024ditto} that are related to chunked autoregression---to the best of our knowledge, these approaches have not been explored for live generation. 

\section{Limitations}
\label{sec:limitations}
Both Magenta RT and Lyria RT present known limitations. Since the models operate on two-second chunks, user inputs for the style prompt may take two or more seconds to influence the musical output. Due to the maximum audio context window of ten seconds, the models are also unable to directly reference music that has been output further into the past. While this is sufficient to create melodies, rhythms, and chord progressions, it does not allow to automatically create longer-term song structures.

\section{Additional Methodological and Training Details}\label{sec:additional_details}

\subsection{SpectroStream}
\label{sec:spectrostream}

Here we adopt the recently proposed SpectroStream  codec~\cite{li2025spectrostream}, a full-band multi-channel neural audio codec based on residual vector quantization (RVQ)~\citep{Zeghidour2021}. 
Similarly to its predecessor SoundStream \cite{Zeghidour2021}, SpectroStream is trained using a combination of adversarial and reconstruction losses. 
Unlike SoundStream, SpectroStream models audio in the time-frequency domain and adopts a delayed fusion mechanism, which together allow for high-fidelity audio. 
Specifically, SpectroStream operates on full-bandwidth stereo music at high sample rate ($f_s = 48$kHz). 
Relative to other discrete codecs like the Descript Audio Codec~\cite{kumar2023high}, 
we train SpectroStream with a relatively slow framerate ($f_k = 25Hz$) and deeper residual quantizers ($d_c = 64$), consistent with recommendations from~\cite{wu2025codesign}. 
Using $10$-bit codebooks ($|\mathbb{V}_c| = 1024$), this induces an overall bandwidth of $16$kbps. 
With the goal of facilitating streaming generation via an LM, 
we reduce the bitrate for generative modeling to $4$kbps by generating only the first $16$ RVQ levels (coarse and medium from~\Cref{fig:magenta_rt_global}), 
inducing a throughput target for live generation of $400$ tokens per second. 

\section{Lyria RT and Advanced Controls}
\label{app:advanced_controls}
\input{advanced_controls}

\section{Audio Injection}
\label{app:audio_injection_details}

\begin{figure}
    \centering
    \includegraphics[width=0.8\linewidth]{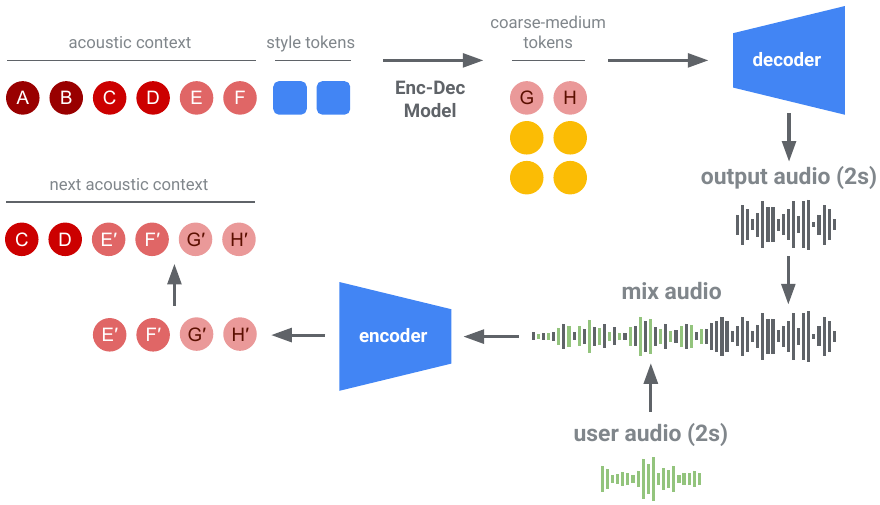}
    \caption{Steering with a live audio stream using \emph{audio injection}. At each inference step, user input audio is mixed with model output audio, and the mix is encoded as coarse SpectroStream tokens. These context tokens are passed as the input for the next inference step, so the model predicts the continuation following its own output \emph{plus} the user's input.}
    \label{fig:audio_injection}
\end{figure}

To allow the user to continuously steer generation via a live audio stream, we propose an audio steering mechanism that we call \textit{audio injection}. At each generation step, we mix the user’s input audio with the model’s output, tokenize the resulting mix, and feed this as the model’s context for the next generation. This is illustrated in Figure~\ref{fig:audio_injection}. Note, user audio is never played back directly. Rather, the model predicts the continuation of a past context that \emph{includes} the user audio. Depending on the specific audio injected and the target style, the model may ``choose'' to repeat or transform the user audio, or may be influenced by its features (dynamics, melody, harmony).

To better control the influence of the user audio on the model output, we use classifier free guidance \cite{ho2021classifierfree}. Specifically, we perform inference on both a positive example where the audio context \emph{includes} the mixed-in user input, as well as a negative example where the audio context consists of only the model’s output. The final logits used for sampling are calculated as a linear combination of the positive and negative logits, where higher guidance weight ($w$) pushes towards the positive conditioning: $(1 + w)\cdot\text{Logits}_{\text{pos}} - w\cdot\text{Logits}_{\text{neg}}$

\paragraph{Live audio prompt} To further increase the effect of the user input, we add a ``live'' audio prompt that periodically updates based on the MusicCoCa audio embedding of the most recent user input audio. Increasing the weight of this prompt within overall prompt mixture has the effect of steering the model to output music in a style that is consistent with the user audio (e.g., containing guitar, if the user is playing guitar).

One challenge to audio injection is that there is unavoidable latency between the input and output streams. In our testing setup, we observed a delay of several seconds between the input and output. Thus, if the user is performing \emph{along with} the model, we will only have around $7$ seconds of input audio ready to mix with the model’s most recent $10$ seconds of output, to feed as the next context.

We consider two solutions to this issue, which are suited to two different interaction styles: ``free'' mode, and ``looper'' mode. In free mode, the user’s input audio is mixed directly with the ``concurrent'' output (i.e.~matching what the user heard while they were performing), but no input is mixed into the final stretch of context. In looper mode, we assume the music follows a ``looping'' structure at a known tempo, and mix input from the \emph{previous} loop into the current context, filling the entire context window with input audio. We discuss these approaches and their tradeoffs below.

\paragraph{``Free'' mode} We mix the user’s input in at its original timing (aligning with what the user heard when they performed it), and mix in silence for that portion of the context for which the streamer has not yet received the inputs. Since the inputs are mixed at their logical location, this control mechanism is intuitive, and the streams can be mixed without knowledge of the tempo or song structure. However, one disadvantage is the user’s contribution will suddenly drop to silence for the final portion of the model’s context. This ``cutting out'' can weaken the effect of the conditioning, as the most recent context tends to have the largest effect on generation. It may also lead to unwanted artifacts, since the user’s audio may be clipped to silence mid-phrase.\footnote{We can soften these artifacts by fading the input to silence. However, the fade out may still be unexpected, and doesn’t correspond to the user’s true performance.}

\paragraph{``Looper'' mode} For music with fixed tempo and a relatively short ``looping'' structure (e.g., a repeating 4-bar chord progression), we can overcome the latency issue by mixing in user audio from the \emph{previous} loop, as opposed to the current one. This interaction is similar to a ``looper'' pedal (where audio is looped to build layered composition); however instead of playing the past audio back verbatim, we mix it into the model’s context window, where it may influence the model’s continuation. Compared to ``free'' mode, this has the advantage of allowing us to fill the \emph{entire} context window with user audio, which we find increases the likelihood of the model being influenced by that audio.

However there are several downsides to ``looper'' mode. First, the control is less intuitive than ``free'' mode, as the effect of the user input is delayed by one loop. This requires the user to plan ahead, and also limits what musical forms can be expressed. Second, the user needs to specify the loop length (e.g., $8$ beats at $120$ BPM), and we need to guarantee the model output actually respects this chosen tempo and loop length. This can be achieved by tempo conditioning (Section \ref{sec:descriptor_conditioning}), or by adjusting the loop length in real time to match the model output. However if the model ever drifts from the set tempo, the mixed user input will be misaligned with the model output, which can lead to unexpected results.

Throughout development, we prototyped audio injection with musicians of different backgrounds, including instrumentalists (guitar, piano, drums), a producer, and a live electronic DJ\@. User responses were varied, with some finding it inspiring, and others finding it too unpredictable.

\section{User Study}
\label{app:user_study}
\input{user_study.tex}

\newpage
\section{Prompt Transitions}
\label{sec:prompt_transition_list}
Prompts used for the ``Prompt Transition'' eval in Section~\ref{sec:transition_evals}. Transitions are linear interpolations of text embeddings every 10 seconds over 60 seconds (i.e., [$0.0$,\,$1.0$], [$0.2$,\,$0.8$], [$0.4$,\,$0.6$], [$0.6$,\,$0.4$], [$0.8$,\,$0.2$], [$1.0$,\,$0.0$]).

\noindent
\begin{tabular}{l@{\hspace{4em}}l}

Accordion $\rightarrow$ Ambient & Accordion $\rightarrow$ Dirty Synths \\
Accordion $\rightarrow$ Minimal Techno & Afrobeat $\rightarrow$ Synthpop \\
Alternative Country $\rightarrow$ Dirty Synths & Alto Saxophone $\rightarrow$ Dulcimer \\
Ambient $\rightarrow$ Gypsy Jazz & American Folk $\rightarrow$ Afrobeat \\
Balalaika Ensemble $\rightarrow$ Fuzz Guitar & Banjo $\rightarrow$ Jamaican Dub \\
Baroque $\rightarrow$ Djembe & Baroque $\rightarrow$ West Coast Hip Hop \\
Bass Clarinet $\rightarrow$ Orchestral Score & Bassoon $\rightarrow$ Classic Rock \\
Blues Rock $\rightarrow$ Balalaika Ensemble & Blues Rock $\rightarrow$ Glitch Hop \\
Blues Rock $\rightarrow$ Marching Band & Bongos $\rightarrow$ Lute \\
Bongos $\rightarrow$ Viola Ensemble & Bouzouki $\rightarrow$ Indie Folk \\
Breakbeat $\rightarrow$ Bongos & Breakbeat $\rightarrow$ Synthpop \\
Cavaquinho $\rightarrow$ Flamenco Guitar & Cello $\rightarrow$ Classic Rock \\
Charango $\rightarrow$ Guitar & Clavichord $\rightarrow$ Precision Bass \\
Congo Drums $\rightarrow$ Vaporwave & Contemporary R\&B $\rightarrow$ Harpsichord \\
Contemporary R\&B $\rightarrow$ Pop Punk & Country $\rightarrow$ Breakbeat \\
Country $\rightarrow$ Renaissance Music & Delta Blues $\rightarrow$ Smooth Pianos \\
Didgeridoo $\rightarrow$ Charango & Didgeridoo $\rightarrow$ Kalimba \\
Dirty Synths $\rightarrow$ Cavaquinho & Dirty Synths $\rightarrow$ Ukulele \\
Disco Funk $\rightarrow$ Hard Rock & Djembe $\rightarrow$ Dirty Synths \\
Djembe $\rightarrow$ Psychedelic & Doo Wop $\rightarrow$ Bossa Nova \\
Doo Wop $\rightarrow$ Industrial Rock & Doo Wop $\rightarrow$ Tango \\
Drum \& Bass $\rightarrow$ Gypsy Jazz & Drum \& Bass $\rightarrow$ Harp \\
Drum \& Bass $\rightarrow$ Trance & Electro Swing $\rightarrow$ Lute \\
Electro Swing $\rightarrow$ Pipa & Erhu $\rightarrow$ Mandolin \\
Fiddle $\rightarrow$ Hard Rock & Fiddle $\rightarrow$ Surf Rock \\
Flamenco Guitar $\rightarrow$ Bass Clarinet & Funk Metal $\rightarrow$ Pipa \\
Funky $\rightarrow$ Koto & Fuzz Guitar $\rightarrow$ Soprano Saxophone \\
Fuzz Guitar $\rightarrow$ Synth Pads & Fuzz Guitar $\rightarrow$ TR-909 Drum Machine \\
Fuzz Guitar $\rightarrow$ Trance & Glitch Hop $\rightarrow$ Synthpop \\
Glockenspiel $\rightarrow$ Clavichord & Hang Drum $\rightarrow$ Jamaican Dub \\
Hang Drum $\rightarrow$ Steel Drum & Hard Bop Jazz $\rightarrow$ Tango \\
Hard Rock $\rightarrow$ Balalaika Ensemble & Hard Rock $\rightarrow$ Hurdy-gurdy \\
Harp $\rightarrow$ R\&B (Rhythm and Blues) & Harpsichord $\rightarrow$ Congo Drums \\
Harpsichord $\rightarrow$ Koto & Harpsichord $\rightarrow$ Trance \\
Heavy Metal $\rightarrow$ Chiptune & Heavy Metal $\rightarrow$ Jamaican Dub \\
Hurdy-gurdy $\rightarrow$ Cavaquinho & Hurdy-gurdy $\rightarrow$ Classic Rock \\
Hurdy-gurdy $\rightarrow$ Gypsy Jazz & Hyperpop $\rightarrow$ Bongos \\
Indian Classical $\rightarrow$ Tuba & Indie Pop $\rightarrow$ Ska \\
Industrial Rock $\rightarrow$ Accordion & K-Pop $\rightarrow$ Soprano Saxophone \\
Klezmer $\rightarrow$ Glockenspiel & Klezmer $\rightarrow$ Hang Drum \\
Klezmer $\rightarrow$ Tuba & Latin Jazz $\rightarrow$ Erhu \\
LinnDrum $\rightarrow$ Synth Pads & Lo-Fi Hip Hop $\rightarrow$ Bagpipes \\
Lute $\rightarrow$ Balalaika Ensemble & Lyre $\rightarrow$ Latin Jazz \\
Lyre $\rightarrow$ Techno & Mandolin $\rightarrow$ Synth Pads \\
Marching Band $\rightarrow$ Chamber Music & Mbira $\rightarrow$ Hard Bop Jazz \\
Neo-Soul $\rightarrow$ Marimba & Neo-Soul $\rightarrow$ Psychedelic Rock \\
Orchestral Score $\rightarrow$ Mellotron & Piano Ballad $\rightarrow$ Garage Rock \\
Piano Ballad $\rightarrow$ LinnDrum & Pipa $\rightarrow$ Chiptune \\
Pop Punk $\rightarrow$ Ambient & Pop Punk $\rightarrow$ Baroque \\
Pop Punk $\rightarrow$ Hurdy-gurdy & Post-Punk $\rightarrow$ Chillout \\
Precision Bass $\rightarrow$ Accordion & Progressive House $\rightarrow$ Irish Folk \\
\end{tabular}

\begin{tabular}{l@{\hspace{4em}}l}
Progressive House $\rightarrow$ Mariachi & Psychedelic $\rightarrow$ Moombahton \\
Psychedelic Rock $\rightarrow$ Sitar & R\&B (Rhythm and Blues) $\rightarrow$ Mariachi \\
Renaissance Music $\rightarrow$ Warm Acoustic Guitar & Sarod $\rightarrow$ Afrobeat \\
Sarod $\rightarrow$ Funk Drums & Sarod $\rightarrow$ Smooth Pianos \\
Sitar $\rightarrow$ Marimba & Ska $\rightarrow$ Dubstep \\
Ska $\rightarrow$ Fiddle & Smooth Pianos $\rightarrow$ Garage Rock \\
Steel Drum $\rightarrow$ Post Rock & Surf Rock $\rightarrow$ Fiddle \\
Surf Rock $\rightarrow$ Gypsy Jazz & Synth Pads $\rightarrow$ Mellotron \\
Thrash Metal $\rightarrow$ Tabla & Trance $\rightarrow$ Djembe \\
Vaporwave $\rightarrow$ Ragtime Piano & Warm Acoustic Guitar $\rightarrow$ Indie Electronic \\
Warm Acoustic Guitar $\rightarrow$ Klezmer & Warm Acoustic Guitar $\rightarrow$ LinnDrum \\
West Coast Hip Hop $\rightarrow$ Indian Classical & Woodwinds $\rightarrow$ Accordion \\
Woodwinds $\rightarrow$ Psychedelic Rock & Zither $\rightarrow$ Dreamy \\
\end{tabular}

\end{document}

%% file: results.tex
Live music models enable new types of interaction that are best assessed via direct human evaluation and extensive use.
As such, many of our design choices were validated through play testing by team members and partner musicians, optimizing for how expressive and engaging the resulting instrument feels. A formalization of this interactive evaluation is presented in our user study in Appendix~\ref{app:user_study}.

We complement this subjective measure with more constrained experiments to examine the effects of our controls and provide comparisons to existing models where possible. We focus these experiments on assessing core capabilities that are common to both live and offline music audio generation models, such as audio quality and adherence to text conditioning (Section~\ref{sec:comparison_to_prior_work}), alongside others that are unique to our live music models, such as the ability to generate musical transitions following changes in the conditioning signal (Section~\ref{sec:transition_evals}). In addition to our evaluation, we also note that, at the time of writing, Magenta RT is ranked as the top open-weights model on the Music Arena leaderboard \cite{kim2025musicarena}, based on over 1k real-world user preferences.

\subsection{Experimental Set-up} \label{sec:experimental_setup}
\paragraph{Training}
We pretrain the LM at Base (220M parameters) and Large (770M parameters) size. The dataset comprises around $190{,}000$ hours of primarily instrumental stock music sourced from various providers. Each training example consists of a $12$-second audio clip randomly sampled from the raw data, structured as $10$-second context tokens and $2$-second target tokens. MusicCoCa tokens are derived from the target audio, of which the first $6$ RVQ levels are used for training. To mitigate the cold-start issue in streaming, we replace early context tokens with variable-length padding tokens.

Each model is trained for $1.86$ million steps using the Adafactor optimizer with batch size of $512$ and an inverse square root learning rate schedule with $10{,}000$ warmup steps. We use TPU-v6e (Trillium) hardware, with $128$ chips for the Base model and $256$ chips for the Large.

\paragraph{Sampling parameters}
For inference, prompts (text or audio) are embedded and tokenized by MusicCoca using the first $6$ RVQ levels to match training. We sample with classifier-free guidance (CFG) \cite{ho2021classifierfree, kreuk2023audiogentextuallyguidedaudio, sanchez2023staytopicclassifierfreeguidance}, using a temperature of $1.3$, a top-K of $40$, and a CFG weight of $5.0$.

\subsection{Results}

\begin{table}[]
\caption{Instrumental music generation results on the Song Describer Dataset \cite{manco2023thesong}. We compare to open models, using a fixed length of $47$s for all samples, though our models are capable of arbitrary-length generation. For all prior models, we report results from \cite{evans2025stable}.}
\resizebox{\columnwidth}{!}{%
\centering
\begin{tabular}{lcccccc}
\toprule
\textbf{Model} & \textbf{Live} & \textbf{Sample rate}  & \textbf{Params} & \textbf{FD}$_{openl3} \downarrow$ & \textbf{KL}$_{passt} \downarrow$ & \textbf{CLAP}$_{score }
\uparrow$           \\
\midrule
{Magenta RealTime}  & \checkmark  & 48 kHz & 760M &\textbf{72.14} &	\textbf{0.47} & 0.35  \\
Stable Audio Open \cite{evans2025stable} & \ding{55} & 44.1 kHz   & 1.1B & 96.51 & 0.55 & \textbf{0.41}   \\ 
MusicGen-stereo-large \cite{copet2023simple} & \ding{55} &  32 kHz & 3.3B & 190.47   &     0.52    &   0.31     \\
\bottomrule
\end{tabular}
}
\vspace{-4mm}
\label{tab:sdd_metrics}
\end{table}

\subsubsection{Audio quality and adherence to fixed text prompts}
\label{sec:comparison_to_prior_work}
In this section, we compare Magenta RT to prior work under the offline text-to-music generation setting, where we keep the text conditioning fixed and sequentially generate chunks of audio up to a target length of $47$ seconds. While our model can generate audio of arbitrary length, we choose this duration for a fair comparison between models. Following the evaluation protocol in \cite{evans2025stable}, we then assess the quality and text adherence of the resulting generations using three established metrics, the Fréchet Distance based on OpenL3 embeddings \cite{cramer2019look} (FD\textsubscript{openl3}), the Kullback–Leibler divergence (KL\textsubscript{passt}) and CLAP\textsubscript{score}.

We show the results in Table~\ref{tab:sdd_metrics}. Magenta RT has the lowest FD\textsubscript{openl3} and KL\textsubscript{passt} scores, indicating that the generated audio is plausible and closely matches the eval reference audio, including at the level of semantic correspondence \cite{evans2025stable}. The CLAP\textsubscript{score} measures how well generated audio adheres to the specified text prompt. On this metric, Magenta RT scores between the other two models. The higher score of Stable Audio Open may be related to the fact that their model uses CLAP embeddings during training, as opposed to our model which trains using MusicCoCa.

\subsubsection{Generating musical transitions}
\label{sec:transition_evals}

\begin{figure}
    \centering
    \includegraphics[width=0.8\linewidth]{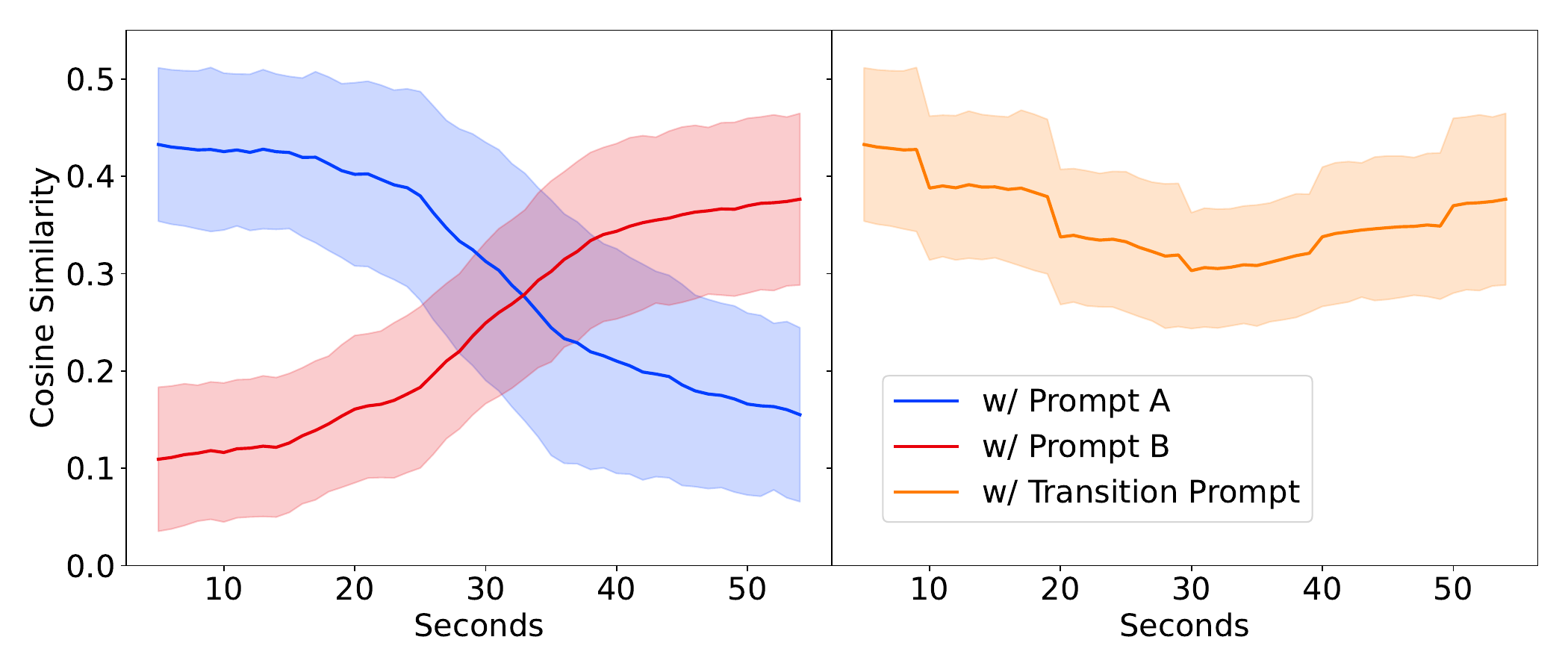}
    \vspace{-2mm}
    \caption{Prompt transition evaluation.
    Over $60$s, we transition from embeddings of text prompt A to B by stepwise linear interpolation. Left: Cosine similarity compared to the initial (blue) and final (red) text embedding. Right: Cosine similarity to the interpolation between text embeddings provided to the model. In both plots, lines indicate the mean and shaded regions the standard deviation.}
    \label{fig:transition_similarity}
    \vspace{-4mm}
\end{figure}

Live music models unlock the capability of responding dynamically to user inputs. To test this, we create a \emph{prompt transition} evaluation. We pick pairs of text prompts and task the model with creating musical transitions between them. We linearly interpolate between MusicCoCa text embeddings of the start and end prompt over $60$ seconds ($6$ steps, $10$ seconds$/$step), and use this as style conditioning. We then measure cosine similarity between the audio embedding of the outputs and the conditioning embedding at that time. We perform transitions for $128$ prompt pairs (Appendix~\ref{sec:prompt_transition_list}).

As seen in Figure~\ref{fig:transition_similarity} Magenta RT outputs maintain strong similarity to the target embedding throughout the transition (right panel), and effectively transitions from the initial to final prompt (left panel). The audio context conditioning lead to smooth transitions that blend styles by preserving elements of the initial prompts. While this leads to lower similarity at end of transition and a slight dip in mid-transition similarity, it also lets the music continuously evolve in a smooth and coherent way, and makes the time history of prompts an important and expressive part of performance.

%% file: controls.tex
\subsection{Style Conditioning via Text and Audio}
\label{sec:style_conditioning}
\looseness=-1
During inference, we can create a target conditioning vector $\mathbf{c}$ by computing a weighted average of the MusicCoCa embeddings corresponding to $N$ control prompts, provided as either text or audio $\mathbf{c} = \sum_{i=1}^N w_i M(\mathbf{c_i}) / \sum_i w_i$, where $w_i$ controls the weight of each prompt.
One advantage of using MusicCoCa embeddings instead of attending to text captions is the ability to perform embedding arithmetic to blend styles---e.g., a weighted sum of \texttt{embed}$($``techno''$)$ and \texttt{embed}$($``flute''$)$ gives a good approximation of \texttt{embed}$($``techno flute''$)$---while also controlling the relative influence of each concept. 

Beyond this, a shared audio-text embedding space for conditioning allows for the use of audio prompts as a more direct way of achieving a specific musical style or instrumentation that may be difficult to express via text. Since audio prompts more closely match the training setting, where style conditioning is obtained from the target audio itself, this type of prompting is also expected to be more effective. 
Furthermore, we are able to mix multiple audio prompts to achieve interpolations of different styles, or mix combinations of text and audio prompts. 
Overall, this type of conditioning offers control over high-level characteristics such as genre, style, instrumentation, mood.

\subsection{Audio Injection}\label{sec:audio_injection}
To allow users to continuously steer generation via a live audio stream, we propose an audio steering mechanism we call \textit{audio injection}. At each generation step, we mix the user’s input audio with the model’s output, tokenize the resulting mix, and feed this as the context for the next generation. This is illustrated in Figure~\ref{fig:audio_injection}. Note, user audio is never played back directly. Rather, the model predicts the continuation of a past context that \emph{includes} the user audio. Depending on the specific audio injected and the target style, the model may ``choose'' to repeat or transform the user audio, or may be influenced by its features (dynamics, melody, harmony).  See Appendix~\ref{app:audio_injection_details} for details.

%% file: contributions.tex
\section*{Contributions and Acknowledgments}
\label{sec:contributions}

Within each \textbf{Bolded Category} of contribution type, contributors are listed alphabetically.
\vspace{1em}

\noindent
\begin{minipage}[t]{0.48\textwidth}

\subsection*{Tech Leads}
Adam Roberts \\
Chris Donahue \\
Kehang Han

\subsection*{Core Contributors}
Antoine Caillon \\
Brian McWilliams \\
Cassie Tarakajian \\
Ian Simon \\
Ilaria Manco \\
Jesse Engel \\
Noah Constant \\
Timo I. Denk \\
Yunpeng Li

\subsection*{Contributors}
Alberto Lalama \\
Andrea Agostinelli \\
Cheng-Zhi Anna Huang \\
Ethan Manilow \\
George Brower \\
Hakan Erdogan \\
Heidi Lei \\
Itai Rolnick \\
Ivan Grishchenko \\
Manu Orsini \\
Matej Kastelic \\
Mauricio Zuluaga \\
Mauro Verzetti \\
Michael Dooley \\
Ondrej Skopek \\
Rafael Ferrer \\
Savvas Petridis \\
Zalán Borsos
\end{minipage}
\hfill
\begin{minipage}[t]{0.48\textwidth}
\subsection*{Lyria RealTime API}
Doug Fritz \\
Ivan Solovyev \\
Jingjing Xie \\
Matthew Tang \\
Olivier Lacombe \\
Peter Morgan

\subsection*{Magenta RealTime Release}
Gus Martins \\
Paige Bailey \\
Omar Sanseviero \\
Tris Warkentin

\subsection*{Product Management}
Jeff Chang \\
Hema Manickavasagam \\
Myriam Hamed Torres

\subsection*{Legal}
Austin Tarango \\
Phoebe Kirk

\subsection*{Program Management}
DY Kim \\
Mahyar Bordbar \\
Moon Park

\subsection*{Executive Sponsors}
Aäron van den Oord \\
Douglas Eck \\
Eli Collins \\
Jason Baldridge \\
Tom Hume
\end{minipage}

\subsection*{Acknowledgements}
Many thanks to the entire Lyria team for their support and feedback.

Special thanks and acknowledgment to Alex Tudor, Arathi Sethumadhavan, Arturas Lapinskas, Ashu Desai, Beat Gfeller, Cătălina Cangea, Chris Deaner, Christian Frank, Colin McArdell, Damien Vincent, Eleni Shaw, Julian Salazar, Kazuya Kawakami, Marco Tagliasacchi, Matt Sharifi, Michael Chang, Reed Enger, RJ Skerry-Ryan, Ron Weiss, Sander Dieleman, Sertan Girgin, Tancred Lindholm, Tobenna Peter Igwe, Victor Ungureanu, and Yotam Mann.

Additional thanks to Chris Reardon, Mira Lane, Koray Kavukcuoglu, and Demis Hassabis for their insightful guidance and support throughout the research process. 

MusicFX DJ was the first deployment of Lyria RealTime and developed in collaboration with our partners from Google Labs including Obed Appiah-Agyeman, Tahj Atkinson, Carlie de Boer, Phillip Campion, Sai Kiran Gorthi, Kelly Lau-Kee, Elias Roman, Noah Semus, Trond Wuellner, Kristin Yim, and Jamie Zyskowski. We give our deepest thanks to Jacob Collier, Ben Bloomberg, and Fran Haincourt for their valuable feedback throughout the development process.

We also acknowledge the many other individuals who contributed across Google DeepMind and Alphabet, including our partners at Envisioning Studio and YouTube.

%% file: advanced_controls.tex
Lyria RT shares the same general architecture as Magenta RT, but with additional controls (see comparison in Table~\ref{tab:compare_mrt_lrt}). Here we detail how those controls are provided as musical descriptors (\ref{sec:descriptor_conditioning}), how they are steered through self-conditioning and control priors (\ref{sec:self-conditioning}), and lastly how the style embeddings are further guided through latent constraints (\ref{sec:mulan_mlp}).

\subsection{Descriptor-based Conditioning}
\label{sec:descriptor_conditioning}

As seen in Section~\ref{sec:style_conditioning}, contrastive audio-text embeddings obtained from models such as MusicCoCa and MuLan effectively enable control over the overall style of generated musical signals. They are not designed, however, to control fine-grained musical attributes. This section investigates the incorporation of additional conditioning features extracted from audio signals using Music Information Retrieval (MIR) methods. A full list of advanced controls for Lyria RT is provided in Table~\ref{tab:descriptors}.

\newcommand{\coltitle}[2]{\makecell[c]{\textbf{#1} \\ \textbf{#2}}}

\begin{table}[]
\centering
\begin{tabular}{lcccccc}
\toprule
\textbf{Model} & \textbf{Access} & \coltitle{Style}{Model}& \coltitle{SS RVQ}{Levels} & \coltitle{Refinement}{Model} & \coltitle{Advanced}{Controls} & \coltitle{Latent}{Constraints} \\ 
\midrule
Magenta RT & Open & MusicCoCa & 16 & \ding{55} & \ding{55} & \ding{55} \\
Lyria RT & API & MuLan~\cite{Huang2022} & 64 & \checkmark & \checkmark & \checkmark \\
\bottomrule
\addlinespace
\end{tabular}
\caption{Comparison of Magenta RT and Lyria RT model features.}
\label{tab:compare_mrt_lrt}
\end{table}

\begin{table}[]
\centering
\begin{tabular}{ll}
\toprule
\textbf{Control} & \textbf{Feature Extractor} \\ 
\midrule
Brightness & Log-mel Spectrogram Spectral Centroid and Bandwidth \\
Density & Onset detection \\
Key & Chroma weighted average \\
Tempo & Beat prediction model~\cite{zhao2022beat} \\
Stems On/Off & Stem separation (Bass, Drums, Vocals, Other), threshold on stem loudness \\
\bottomrule
\addlinespace
\end{tabular}
\caption{List of controls used in training Lyria RT and method of musical descriptor feature extraction.}
\label{tab:descriptors}
\end{table}

\paragraph{Temporal conditioning} We aim to provide precise control over the tempo of the generated stream, defined using \textit{beats-per-minute} (BPM). Since we do not have access to BPM annotated audio examples, we use an off-the-shelf beat prediction model \cite{zhao2022beat} to annotate our dataset, yielding an estimate for beat positions and an averaged BPM value for the entire track. We start by conditioning our model on the target BPM value rounded to the nearest integer using cross-attention.

\paragraph{Instrumentation, timbre and harmony} In addition to style steering using MuLan embeddings, we provide finer grained controls for these features. We use a source separation model \cite{kim2021kuielab} to extract the vocals, bass, drums and other stems from our dataset, and use those stems to generate a set of acoustic features to further condition our model. We specifically extract peak loudness, spectral centroid and bandwidth, chromas and transients separately for every stem.

\subsection{Self-conditioning}
\label{sec:self-conditioning}

\paragraph{Predicting conditioning tokens}

\begin{figure}
    \centering
    \includegraphics[width=\linewidth]{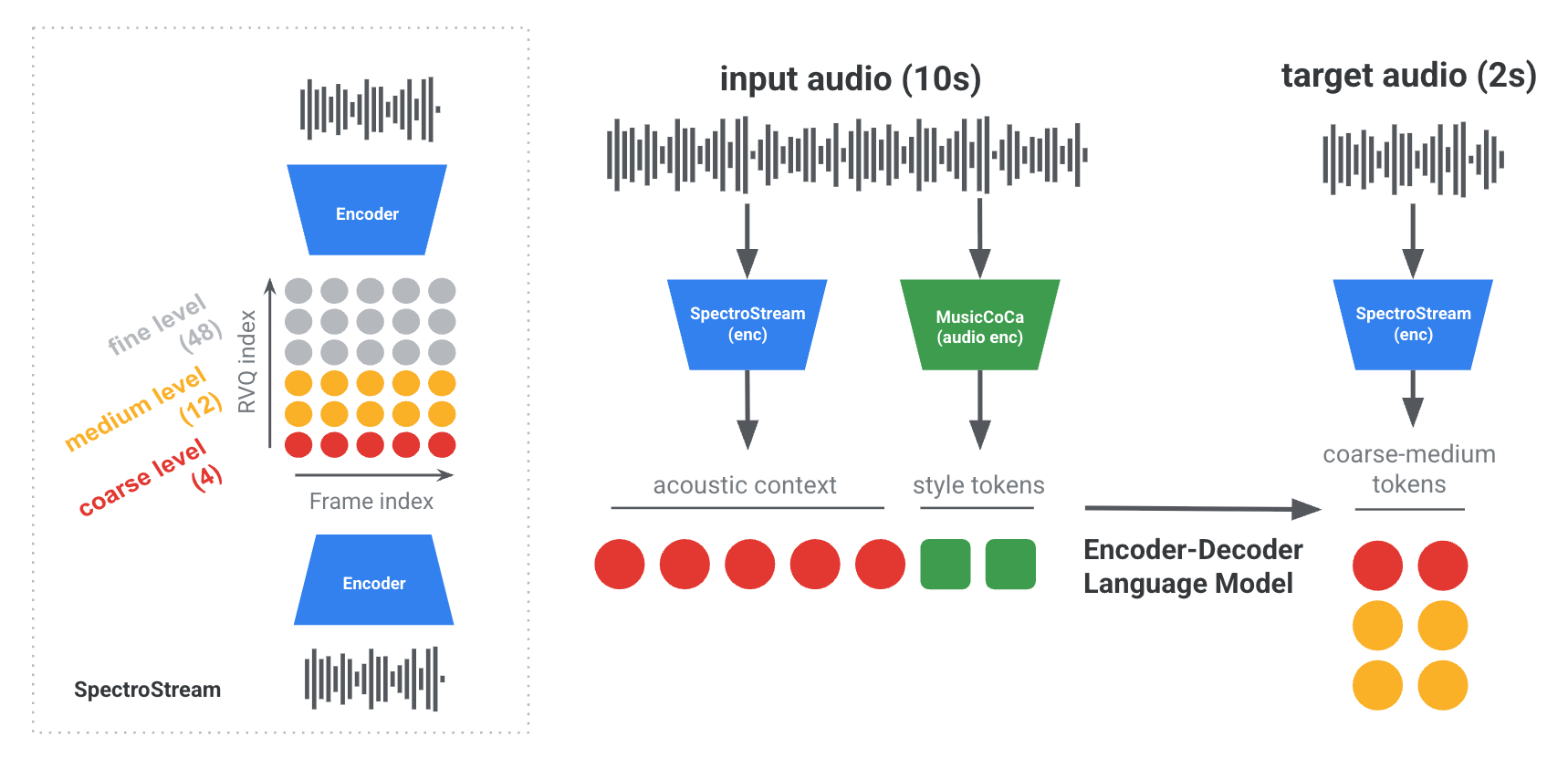}
    \caption{Overall architecture of Magenta RT. Coarse acoustic tokens and quantized style tokens corresponding to 10s of audio context are concatenated and fed to the encoder part of our model. The decoder then predicts coarse and medium acoustic tokens corresponding to the the following 2 seconds.}
    \label{fig:magenta_rt_global}
\end{figure}

\begin{figure}
    \centering
    \includegraphics[width=.9\linewidth]{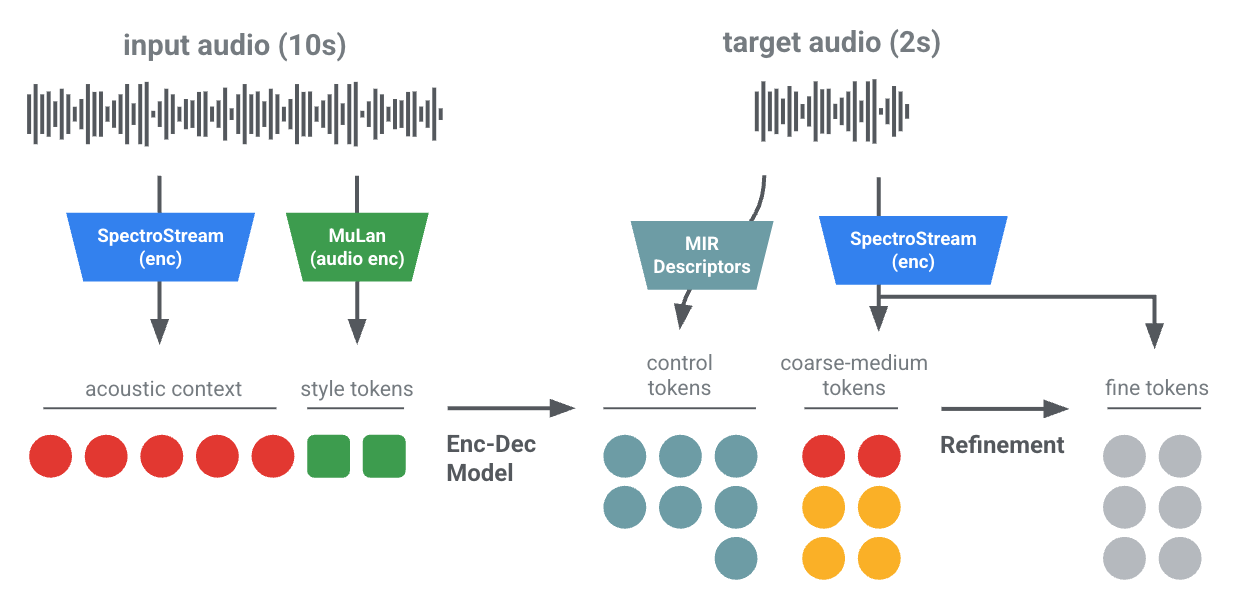}
    \caption{
    Diagram of Lyria RT training / predicting conditioning tokens. Coarse acoustic tokens and quantized MuLan tokens are concatenated and fed to the encoder part of our language model. Control  tokens, including BPM, stem balance, brightness, density and chromas (see Section~\ref{sec:descriptor_conditioning}) are predicted first by the decoder, followed by coarse and medium level acoustic tokens. Finally, a small refinement model predicts the fine-scale acoustic tokens as described in Section~\ref{sec:refinement}.}
    \label{fig:self-conditioned}
\end{figure}

Given an acoustic token sequence $\mathbf x$ and a conditioning token sequence $\mathbf c$, it is common practice to learn the conditional distribution $p(\mathbf x | \mathbf c)$ in a supervised fashion, and expose the conditioning tokens to the user during inference to make the generation controllable. This approach has several drawbacks:

\begin{enumerate}
    \item User defined conditioning might be out of distribution, especially when conditioning tokens are dependent on each other (e.g., natural temporal evolution of a conditioning signal). 
    \item Unconditional generation can be achieved through training-time conditioning dropout, however the resulting samples are empirically worse than those from a model trained unconditionally from scratch.
\end{enumerate}

To address these issues, we introduce \textit{self-conditioning} where we aim at learning the joint distribution $p(\mathbf x, \mathbf c) = p(\mathbf x | \mathbf c) p(\mathbf c)$. This can be achieved by prefixing the acoustic tokens with the conditioning tokens, and train the model using a causal mask, as seen in Figure~\ref{fig:self-conditioned}.

Intuitively, this allows to use the resulting model either unconditionally, by first sampling the conditioning tokens and then sampling from the conditional distribution, or let the user override some or all of the conditioning tokens to gain some control over the generation. While this helps bridge the quality gap with unconditional models, this methods expects the user to provide conditioning tokens aligned with the underlying prior distribution $p(\mathbf c)$. This is a reasonable assumption for scalar conditions like tempo, but is significantly harder for complex conditions such as stems loudness or chromas.

\paragraph{Control priors} 

\begin{figure}
    \centering
    \includegraphics[width=.8\linewidth]{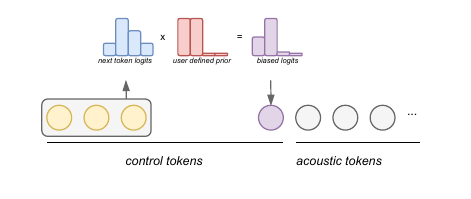}
    \caption{Control priors for self-conditioning. The predicted logits (likelihood) for the control tokens are shifted by a user defined prior dictated by the control values. These are combined to give the final posterior logits that are used for sampling, and steer the model outputs in the direction of the user controls.}
    \label{fig:control_priors}
\end{figure}

We leverage the fact that we already have an estimation of the distribution underlying the conditioning tokens $p(\mathbf c)$ to implement \textit{soft controls}. Soft controls are implemented through the definition of a prior distribution over the conditioning tokens that is used to steer the sampling process. In practice, we map simple controls to categorical priors that we combine with the next token predicted logits to steer the sampling process, as shown in Figure \ref{fig:control_priors}.

\subsection{Constraining Style Embeddings}
\label{sec:mulan_mlp}

As mentioned in Section~\ref{sec:musiccoca}, we train the language models conditioned on style embeddings computed from the raw audio waveform. This is in contrast to the inference setup where we use text based embeddings. While similar, due to the contrastive training of MuLan, the audio and text embedding spaces are not completely overlapping. Indeed, early experiments showed that a two layer MLP is sufficient to classify embeddings as text-based or audio-based with $>90\%$ accuracy.
In practice, we notice a non-negligible drop in audio quality when predicting from text based embeddings compared to audio embeddings, which is likely due to the mismatch between both embedding spaces.

Furthermore, biases exist between specific text genres and audio recording quality. 
To address these issues, impose a \textit{latent constraint}~\cite{engel2018latent}, by building a dataset comprised of MuLan embeddings from three sources (text, low-quality audio, and high-quality audio) and training a small GAN to transform text based embeddings into high quality audio embeddings using an adversarial setup.

Both the generator and discriminator are a 4-layer MLP. The discriminator is trained to separate ground truth high-quality audio embeddings (i.e. \textit{real) embeddings} from the generator outputs (i.e. \textit{fake} embeddings). The generator is trained to optimize the following two objectives: being classified as \textit{real} by the classifier while staying close to the original base embedding. We use the Hinge GAN objective, and use cosine similarity between input and generated embeddings as a regularization strategy.

\begin{figure}
    \centering
    \includegraphics[width=.9\linewidth]{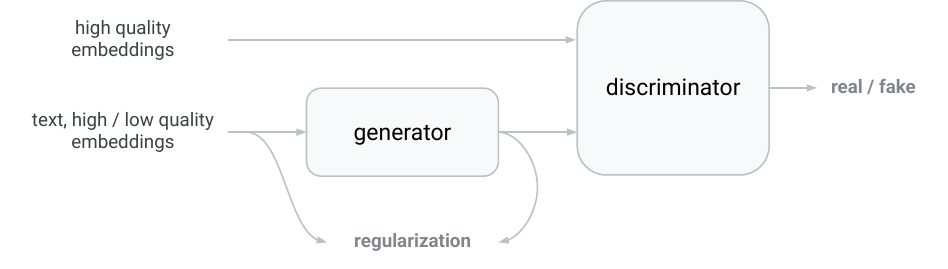}
    \caption{Training the latent constraint model}
    \label{fig:mulan_mlp}
\end{figure}

\subsection{Refinement Model}
\label{sec:refinement}
As seen in Figure~\ref{fig:self-conditioned}, Lyria RT has an additional refinement model to predict 48 additional fine-scale RVQ SpectroStream tokens per a frame. This model is trained separately from the core language model, and is a very small MLP that quickly autoregressively predicts the remaining tokens. This provides a slight bump in audio fidelity at the cost of longer inference. Because Magenta RT is focused towards real-time control on less powerful hardware, we reserve the refinement model for Lyria RT and find it to be a good compromise of audio fidelity and compute efficiency.

%% file: user_study.tex
To understand how live music models impact users’ creative expression and process, we conducted a study with five music enthusiasts. We were primarily interested in probing how the continually streaming nature of the model impacted their state of flow and creative expression as well as their perceived level of control of the model’s generated music.

\begin{figure}
    \centering
    \includegraphics[width=0.7\linewidth]{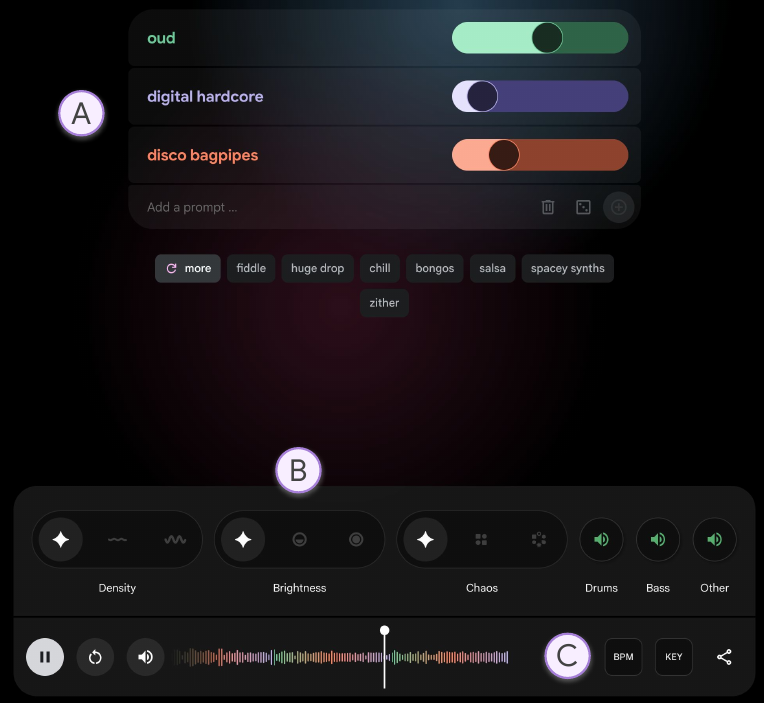}
    \caption{Screenshot of the interface used in our study.}
    \label{fig:user_study_ui}
\end{figure}

\paragraph{User interface}
As part of the study, we used the MusicFX DJ interface developed for the Lyria RT model,\footnote{\url{http://labs.google/musicfx}} which includes the advanced controls described in Section~\ref{sec:descriptor_conditioning}.
The interface, shown in Figure~\ref{fig:user_study_ui}, allows users to input their own natural language prompts (e.g. “oud” or “digital hardcore”), each accompanied by a slider that adjusts its respective influence on the generated music. The system also provides prompt suggestions for inspiration. In addition to steering the model through these prompts, users are also given three high-level controls: density, which adjusts how busy the music should be; brightness, which adjusts the presence of higher frequency sounds; and chaos, which controls how unpredictable the generated music should be. To begin streaming music, users press play, and as they make adjustments to the prompts or high-level controls, the music adjusts accordingly. Finally, users can adjust the key and BPM of the produced music, as well as clear the music history if they want to start fresh.

\paragraph{Study structure}
The overall structure of the user study was as follows: (1) participants were given a short demo of the user interface, (2) they then interacted with the interface for 20 minutes, while thinking aloud and explaining their thought process, and (3) afterwards, they were interviewed on their experience. During the task, participants were asked to simply explore the user interface, treat it as an “interactive radio” and discover new sounds (Section~\ref{sec:instructions}). Afterwards, the interview consisted of questions that probed their thoughts on the impact of the model’s continually streaming nature and their perceived level of control (Section~\ref{sec:exit_interview}). The total time commitment of the study was 30 minutes.

\subsection{Findings}
\paragraph{Impact of continuous streaming}
The continuous nature of the generated music significantly shaped the participants' creative experience; it gave them something to respond to and improvise with, often drawing comparisons to collaborative and traditional music-making processes. P1 compared interacting with the continual stream to improvising with fellow musicians: “It’s not so different from playing with other people... as you play you react to each other non-verbally to the music. I can see three people playing instruments along with this, steering the model together”. This sentiment was echoed by two other musicians and composers, P4 and P5, who felt the experience mirrored their own composition workflows of creating loops, reacting to them, and incrementally layering additional tracks on software like Ableton. For example, P5 started off with warm acoustic guitar, and then subsequently layered in flute and harp. When asked about her process, P5 explained that the acoustic guitar prompt led to a feeling of “walking in a forest after some rain”, and she added in flute to “mimic the sound of birds chirping”, as well as the harp to further extend the feeling of peace she was building. 

A key element of the live model’s continuous stream that users appreciated was its ability to introduce subtle, ongoing variations, even when they had not changed the prompts. This element of gentle evolution was highly valued by P3, who appreciated the opportunity for serendipitous discovery it provided: “I wait to see if the model will produce something interesting... the model might have settled on something that was originally boring, but then jump to something more exciting”. Echoing this point, P4 explained that the subtle variations were “nice in that it [the music] doesn’t feel repetitive”, but at the same time, he noted that these variations would also make him feel like he had “lost something”, if there was something he wanted to keep. Sometimes a melody or texture he liked would suddenly drop out and he felt he had no way to get it back or reinforce it. 

\paragraph{On control and influencing the model}
In regards to controlling this continuous stream of music, participants felt that the prompts let them broadly steer the model, rather than afford them specific, direct control. P3 felt like he was “guiding and the model is meeting you half way”, while P4 described it as “throwing ingredients in the pot, and the AI is cooking it up”. This perceived level of control, however, changed significantly according to the number of prompts already included. For instance, P1 noted that while the model could be “extremely accurate” with one or two prompts, it became less predictable as more were added. Similarly, P4 felt he had a “good amount of control” when creating an initial sound, but “a lot less control” when trying to make nuanced, real-time adjustments, where additional prompts could suddenly “take over the track” and completely change the sound.

\paragraph{Summary}
Overall, the live model’s continuous output fostered a collaborative and improvisational creative process, allowing users to react to and build upon the AI's output as they would with a fellow musician. While participants appreciated the model's evolving variations for sparking serendipitous discovery, they ultimately felt more like a guide than a director, possessing the ability to broadly steer the music but lacking the precise control needed to refine it or prevent the loss of desired elements.

\subsection{Participant Instructions}\label{sec:instructions}
We report below the instructions given to participants at the beginning of each session:

\emph{Introduction}:
Hi, thanks for taking the time to participate in this study. During this study, we’ll be creating some sounds with a live, generative music tool. 
We’ll spend about 15 minutes using the tool, where you’ll be asked to think aloud and explain your thought process as you use it. And afterwards we’ll have a 5-10 minute exit interview on your experience.
Any questions so far?

\emph{Task}:
Please point your browser to:  https://labs.google/fx/tools/music-fx-dj and share your screen (as well as audio).
For the next 20 minutes we’ll be playing around with this tool. As you try out its features, please think-aloud and describe your thought process and reactions. 
Imagine you’re a composer using this tool to explore new sounds or inspiration for a new project.

\subsection{Exit Interview Questions}\label{sec:exit_interview}
After completing the task, participants were asked the following questions:
    
\emph{Overall experience}: (1) Could you describe your overall experience using the tool? What was surprising, exciting, or frustrating?

\emph{Exploration process}: (1) Describe a moment when you discovered a particularly interesting or unexpected sound or musical direction. What did you do to get there? (2) What was your thought process behind choosing new prompts or adjusting the sliders? (3) How did the continuous nature of the music influence your exploration or creative process?

\emph{Prompt history}: (1) How would you describe how the music changed and evolved over the course of the session? (2) How did you feel the history of your prompts shaped the music the model produced later on?

\emph{Agency \& control}: (1) Could you describe the level of control you felt over the music that was generated? (2) How would you describe your partnership with AI in creating the music you heard?